\documentclass[10pt, onecolumn]{article}

\usepackage{bm,color,bbm}
\usepackage{ulem} 
\usepackage{hyperref, mathtools,graphicx} 
\usepackage{mathtools,graphicx}

\newcommand{\beq}{\begin{equation}}
\newcommand{\eeq}{\end{equation}}
\newcommand{\bqa}{\begin{eqnarray}}
\newcommand{\eqa}{\end{eqnarray}}
\newcommand{\nn}{\nonumber}

\newcommand{\smallfrac}[2]{\mbox{$\frac{#1}{#2}$}}

\newcommand{\half}{\smallfrac{1}{2}}

\newcommand{\id}{\mathbbm{1}}
\newcommand{\h}{{\cal H}}

\definecolor{maroon}{rgb}{0.7,0,0}

\definecolor{ngreen}{rgb}{0.3,0.7,0.3}

\definecolor{golden}{rgb}{0.8,0.6,0.1}

\begin{document}
\title{On two recent proposals for witnessing\\ nonclassical gravity}
\author{
	Michael J. W. Hall\thanks{Centre for Quantum Computation and Communication Technology (Australian Research Council), Centre for Quantum Dynamics, Griffith University,
		Brisbane, QLD 4111, Australia} \thanks{Department of Theoretical Physics, Research School of Physics and Engineering,
		Australian National University, Canberra ACT 0200, Australia} and Marcel Reginatto\thanks{Physikalisch-Technische Bundesanstalt, Bundesallee 100, 38116 Braunschweig, Germany}}
\date{}
\maketitle



\begin{abstract}
	Two very similar proposals have been made recently for witnessing nonclassical features of gravity, by  Bose {\it et al.} and by Marletto and Vedral.
However, while these proposals are asserted to be very general, they are in fact based on a very strong claim: that quantum systems cannot become entangled via a classical intermediary. We point out that the support provided for this claim is only applicable to a very limited class of quantum-classical interaction models, corresponding to Koopman-type dynamics. We show that the claim is also valid for mean-field models, but that it is contradicted by explicit counterexamples based on the configuration-ensemble model. Thus, neither proposal provides a definitive test of nonclassical gravity.
\end{abstract}


\section{Introduction}

\begin{figure}[!h] \label{figure}
	\centering
	\includegraphics[width=0.5\textwidth]{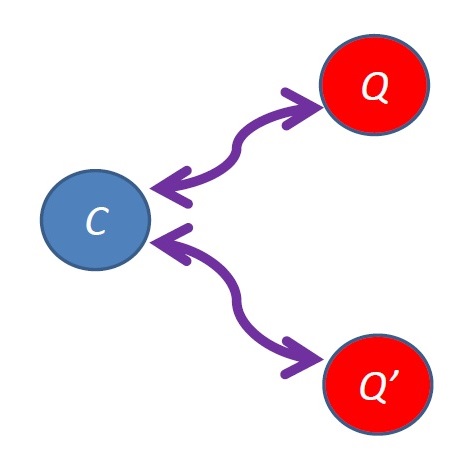}
\caption{Two quantum systems, $Q$ and $Q'$, interact with a classical mediating system $C$. It is claimed in \cite{bose} and~\cite{marl} that such interactions cannot create entanglement between $Q$ and $Q'$. However, we show  that the validity of this claim in fact depends on the model describing the quantum-classical interactions. It follows that observing the creation of entanglement between two quantum masses, as per the experimental proposals in \cite{bose} and~\cite{marl}, does not provide an unambiguous witness of nonclassical gravity. }
\end{figure}

There is a long-running debate as to whether gravity must be quantised (see, e.g.,~\cite{carlip,ark,boughn,dyson} for recent summaries and citations). Very recently, two proposals have been made for settling the matter experimentally~\cite{bose,marl}.  These proposals are very similar, and are based on the same claim: that two quantum systems cannot become entangled via interactions with a classical mediator (see figure~\ref{figure}).

The above claim is asserted to hold very generally. For example Bose {\it et al.}  write  {\it ``this
	entanglement can only result from the exchange of quantum
	mediators''}~\cite{bose},  while Marletto and Vedral write that {\it ``this argument is general and does not rely on any
specific dynamics''}~\cite{marl}. The main point of this paper is to note that the claim {\it does} rely on a specific model of quantum-classical dynamics, and hence does {\it not} provide an unambiguous witness of quantum mediators.

First, in section~\ref{sec:overview}, we emphasise that there are many possible models of quantum-classical interactions, and that these typically have mathematical or physical difficulties associated with them. The particular type of model considered in~\cite{bose} and~\cite{marl} is only one such model, based on formally embedding a classical system into a diagonal basis of some quantum system,  i.e., to describing  quantum-classical interactions via Koopman-type dynamics~\cite{koop,sudar,terno}.  In section~\ref{sec:koop} we confirm the claim made in~\cite{bose, marl} that, for Koopman models, a classical mediator indeed cannot generate entanglement between two quantum systems.  We further show, in section~\ref{sec:mean}, that the claim is also valid for mean-field models of quantum-classical interactions~\cite{boucher}, thus extending the import of the proposed experiments.

However, in section~\ref{sec:config} we demonstrate there are counterexamples to the claim, based on the description of quantum-classical dynamics via the formalism of ensembles on configuration space~\cite{hr2005, hrbook}. In section~\ref{sec:grav} we argue that such counterexamples are expected even for specifically   {\it gravitational} models of quantum-classical interactions.  Results are discussed in section~\ref{sec:disc}, as well as several subtleties.

\section{Brief overview of approaches to\\ quantum-classical interactions}
\label{sec:overview}

Finding a physically consistent approach to modelling interactions between quantum and classical systems is a highly nontrivial task.  Many proposals have been made, which typically have concomitant difficulties of some sort, as we now briefly discuss. The first three classes of models below are of direct relevance to later sections of this paper.

First, the
{\it Koopman} approach relies on modelling the classical system by a set of mutually commuting observables on some Hilbert space, and allowing a unitary interaction with the quantum system.   Formally  embedding a classical system into a quantum system in this way was introduced by Koopman~\cite{koop}, for phase space observables, and its use for describing quantum-classical interactions was pioneered by
Sudarshan and co-workers \cite{sudar, C9_SS1978, C9_SS1979, C9_GSS1979, sudarlatest}.
The models in~\cite{bose} and~\cite{marl} are of the Koopman-type (see section~\ref{sec:koop}). However, we note that Peres and
Terno have shown that Koopman models do not reproduce the correct classical limit for quantum-classical oscillators, and indeed may result in a runaway increase of the  classical oscillator amplitude \cite{terno,C9_terno}.  Further, the simplest such model of classical gravity~\cite{ktm} has already been ruled out by experimental observations~\cite{zych}. 
Diosi et al. have proposed a variation on this approach, in which classical phase space parameters are mapped to a set of coherent states rather than to a set of orthonormal states \cite{diosi}.  However, this variation doe not yield the classical equations of motion in the limit of no interaction, and intrinsically imposes quantum uncertainty relations upon the classical system.

Second, in the {\it mean-field} approach, observables of a classical system appear as parameters in a quantum Hamiltonian operator.  This operator directly specifies the evolution of the quantum system in the usual way, while its average over the quantum degrees of freedom specifies a classical Hamiltonian for the classical parameters \cite{boucher,C9_M1999,zhang, alonso,elze,radon} (see also section~\ref{sec:mean}). Such models are of particular interest in the context of the proposals in \cite{bose} and~\cite{marl}, as they have often been used in semi-classical gravity to model the effects of gravity on quantum systems~\cite{semigrav}. However, while computationally useful as a semiclassical approximation to a fully quantum model, the classical system evolves deterministically.  Thus, the mean-field approach cannot couple any quantum uncertainties into the classical parameters, where such a coupling is required, for example, if measurement and scattering interactions are to lead to a multiplicity of possible outcomes \cite{boucher}.  Nevertheless,
Elze has shown that the mean-field approach  satisfies several basic consistency criteria \cite{elze}. 

Third, the {\it configuration ensemble} approach is based on a very general formalism for describing physical systems in terms of ensembles evolving on a configuration space, with the dynamics specified by an action principle~\cite{hr2005, hrbook}. It encompasses both standard classical and quantum ensembles, and allows for a natural description of interactions between any two ensembles (see section~\ref{sec:config}). The description of generic particle interactions between classical and quantum systems requires the classical component to have a large number of degrees of freedom to avoid nonlocality issues~\cite{ hrbook}, although these issues do not arise for the gravitational case of quantum fields interacting with classical spacetime~\cite{ark,hr2005,hrbook,savage,marcel} (see section~\ref{sec:grav} and Appendix).

Fourth, the
{\it trajectory} approach is based on the deBroglie-Bohm formulation of quantum mechanics, in which quantum systems are described by an ensemble of  trajectories acted on by a `quantum potential' \cite{C9_B1952,C9_B1987,C9_H1993}. Interaction with a classical system is incorporated by modifying the equations of motion for the Bohmian trajectories in various ways \cite{C9_GMB2000,C9_PB2001,C9_BP2004}.  This approach incorporates backreaction on the classical system, and has been found useful for semiclassical calculations in quantum chemistry.  However, like the mean-field approach the classical motion is deterministic, and moreover does not respect energy conservation \cite{C9_S2003,C9_PB2003}.

Finally, various counterexamples and no-go theorems show that other proposed types of quantum-classical interaction lead to at least one of the following problems: negative probabilities; the absence of any backreaction on the classical system from the interaction; or to the loss of the correspondence principle in the classical limit
\cite{terno, boucher, diosi, salcedo, C9_CS1999, sahoo}.

\section{Quantum-classical interactions tested by the proposals
}

 We  consider the general case of two quantum systems $Q$ and $Q'$, with corresponding Hilbert spaces ${\cal H}_Q$ and ${\cal H}_{Q'}$, interacting with a classical system $C$, as depicted in figure~\ref{figure}. In this section we verify the claim in~\cite{bose, marl}, that this interaction cannot create entanglement between $Q$ and $Q'$ for the case of Koopman-type models.  We show that the claim is also valid for mean-field models.

\subsection{Koopman models}
\label{sec:koop}

Marletto and Vedral write that `classical' means {\it ``C has only a single observable''}~\cite{marl}. For a classical point particle, for example, this observable would correspond to the coordinate $(x,k)$ in phase space. They then simply assume that this observable can be mapped to some operator $\hat C$ (or a commuting set of such operators) in some Hilbert space ${\cal H}_C$, and apply the usual theory of quantum interactions to the tensor product $\h_{Q}\otimes \h_{Q'}\otimes \h_C$, under the restriction that all relevant operators are diagonal with respect to $\hat C$. Note,  for example, that this corresponds to representing a classical phase space observable $(x,k)$ by commuting operators $(\hat x,\hat k)$ on a `classical' Hilbert space, as introduced by Koopman~\cite{koop,sudar,terno} (see also section~\ref{sec:overview}).

Marletto and Vedral only discuss their Koopman-type interaction model for the example of a classical bit~\cite{marl} (Koopman models for discrete classical systems  have also been  considered in~\cite{sudarlatest}).  More generally, for spectral decomposition $\hat C=\sum_c c\,\hat \Pi_c$ of $\hat C$, with corresponding projection-valued measure $\{\hat \Pi_c\}$ satisfying $\hat \Pi_c\geq 0$, $\sum_c\hat\Pi_c=\hat \id_C$, and $\hat\Pi_c\hat\Pi_{c'}=0$ for $c\neq c'$, we may take the initial density operator to have the `diagonal' form
\beq
\hat\rho_{QQ'C}= \oplus_c \big[ p(c)\hat\rho_{QQ'}(c)\otimes \hat\rho_C(c) \big] 
\eeq
 with respect to $\hat C$,  where $\hat\rho_C(c)$ is a density operator on the subspace ${\h}_c$ of $\h_C$ corresponding to the unit eigenspace of $\hat \Pi_c$ (thus ${\h_C}=\oplus_c {\h}_c$). Further the final density operator describing $Q$ and $Q'$ has the form
\begin{align}
\hat\rho_{QQ'}^{\rm out} &= {\rm tr}_C[\hat U_{QC}\hat U_{Q'C}\,\hat\rho_{QQ'C}\,\hat U_{QC}^\dagger \hat U_{Q'C}^\dagger] \nn \\
\label{out}
& = \sum_c p(c)\, \hat U_Q(c)\otimes \hat U_{Q'}(c)\,\hat\rho_{QQ'}(c)\, \hat U_Q(c)^\dagger \otimes \hat U_{Q'}(c)^\dagger ,
\end{align}
where $\hat U_{QC}=\oplus_c \left[\hat U_Q(c)\otimes \hat\id_{Q'}\otimes \hat U_C(c)\right]$ and $\hat U_{Q'C}=\oplus_c \left[ \hat\id_Q\otimes \hat U_{Q'}(c)\otimes \hat U'(c)\right]$ are unitary operators describing the respective `diagonal' quantum-classical interactions,  between $Q$ and $C$ and between $Q'$ and $C$, respectively.  Here $\hat \id_Q$ and $\hat \id_{Q'}$ are the identity operators on $\h_Q$ and $\h_{Q'}$, and $\hat U_C(c)$ and $\hat U'_C(c)$ are  unitary operators on ${\h}_c$ (one could replace unitary operations by more general channels; however, this is equivalent to unitary operations on a larger Hilbert space). Since equation~(\ref{out}) is just a convex combination of local unitary operations on $\h_Q\otimes\h_{Q'}$, the quantum-classical interactions cannot increase  entanglement between $Q$ and $Q'$, confirming that Marletto and Vedral's assumptions  imply that no entanglement can be created by such interactions.

In contrast, Bose {\it et al.} simply postulate that the quantum-classical interaction is such that $Q$ and $Q'$ undergo local operations and classical communication (LOCC)~\cite{bose}. Since this is well known not to increase entanglement between quantum systems, they reach the same conclusions as Marletto and Vedral. Noting that LOCC can always be modelled as unitary operations on a sufficiently large Hilbert space, with any classical labels for operations, measurement outcomes and signals represented by a set of mutually commuting operators, their postulate is formally equivalent to assuming that there is a classical mediating system $C$ which can be diagonally embedded into a suitable quantum system.  In this sense, it again corresponds to a Koopman model for quantum-classical interactions. 

Finally, we observe from equation~(\ref{out}) that, in both cases, additionally making a classical measurement on $C$ (corresponding to measuring an observable commuting with $\hat C$ in the Marletto and Vedral approach, and to measuring  local classical labels of operations, outcomes and/or signals in the Bose {\it et al.} approach), cannot create any entanglement between initially separable quantum systems $Q$ and $Q'$, even when postselected on the measurement outcome, since it corresponds to a local classical operation.

\subsection{Mean-field models}
\label{sec:mean}

As noted in section~\ref{sec:overview}, mean-field models are of particular interest in the context of the proposals in \cite{bose} and~\cite{marl} (despite not being considered there), as they have often been used in semi-classical gravity to model the effects of gravity on quantum systems~\cite{semigrav}.

In particle mean-field models the classical system C is represented by phase space coordinates $(x,k)$ and the quantum system by a wave function $|\psi\rangle$, and the dynamics is specified by a Hamiltonian operator $\hat H(x,k)$ that is parameterised by the classical coordinates. In the simplest formulation~\cite{boucher,elze}, the equations of motion are given by the Schr\"odinger equation
\beq i\hbar\frac{d}{dt}|\psi\rangle = \hat H(x,k)|\psi\rangle
\eeq
for $|\psi\rangle$, and Hamilton's equations of motion
\beq
\dot x = \nabla_k \bar H(x,k),\qquad \dot k = -\nabla_x \bar H(x,k)
\eeq
for $x$ and $k$, where $\bar H(x,k):=\langle \psi|\hat H(x,k)|\psi\rangle$ is a classical `mean field' Hamiltonian. Here $\nabla_k$ ($\nabla_x$) denotes the vector of partial derivatives with respect to the components of $k$ ($x$).

The above equations couple the quantum and classical dynamics in a nonlinear manner. Thus, for example,  $|\psi\rangle_0+|\phi\rangle_0$ does not evolve into $|\psi\rangle_t+|\phi\rangle_t$ in general. Hence mean-field models are incompatible with an embedding of the classical system in a quantum system, in contrast to the Koopman approach.  However, we will nevertheless show that, similarly to Koopman models, entanglement cannot be created between two quantum systems via a mean-field model of quantum-classical interactions.

In particular, for two quantum systems $Q$ and $Q'$, consider a coupling to a classical system $C$ as per figure~\ref{figure}, via a mean-field Hamiltonian operator of the form
\begin{align}
\hat H(x,k) &= \hat H_{Q}(x,k)\otimes \hat\id_{Q'} + \hat \id_Q\otimes \hat H_{Q'}(x,k) ,
\end{align}
where $\hat H_{Q}(x,k)$ and $\hat H_{Q'}(x,k)$ are operators on $\h_Q$ and $\h_{Q'}$.
If the initial state of $Q$ and $Q'$ is unentangled, i.e., $|\psi\rangle_0 = |\psi_Q\rangle\otimes |\psi_{Q'}\rangle$, it follows that the state at any later time $t$ has the form
\begin{align}
|\psi\rangle_t &= \left[ {\cal T} e^{-i \int_0^t ds\, \hat H_Q(x_s,k_s)  /\hbar}|\psi_Q\rangle \right]\otimes
\left[  {\cal T} e^{-i \int_0^t ds\, \hat H_{Q'}(x_s,k_s)  /\hbar} |\psi_{Q'}\rangle \right] \nn \\
&=: |\psi_Q\rangle_t\otimes |\psi_{Q'}\rangle_t , \label{fact}
\end{align}
where ${\cal T}$ denotes the time-ordering operator, and the classical phase space trajectory $(x_t,k_t)$ is implicitly determined via the solutions of
\begin{align*}
\dot x_t &=  \nabla_k\left[ {}_t\langle\psi_Q|\hat H_{Q}(x_t,k_t)|\psi_Q\rangle_t +   {}_t\langle\psi_{Q'}|\hat  H_{Q'}(x_t,k_t)|\psi_{Q'}\rangle_t \right],\\
\dot k_t& = -  \nabla_x\left[ {}_t\langle\psi_Q|\hat H_{Q}(x_t,k_t)|\psi_Q\rangle_t +   {}_t\langle\psi_{Q'}|\hat  H_{Q'}(x_t,k_t)|\psi_{Q'}\rangle_t \right] .
\end{align*}
Equation~(\ref{fact}) implies that no entanglement is created between $Q$ and $Q'$. Hence the experimental proposals in \cite{bose} and~\cite{marl} are able to rule out mean-field models, in addition to Koopman models.

\section{Counterexamples: Quantum-classical models not tested by the proposals}
\label{sec:config}

As noted in section~\ref{sec:overview}, another class of models for describing quantum-classical interactions is provided by the formalism of ensembles on configuration space~\cite{hr2005,hrbook}.  This formalism is underpinned by a very simple physical picture: ensembles evolving on a configuration space. The mathematical structure is correspondingly simple (simpler than $C^*$-algebras for example), yet is sufficiently nontrivial to guarantee the existence of, for example, a  dynamical  bracket for observables, thermal ensembles, weak values, and a generalised Ehrenfest theorem. It also forms a natural starting platform for several axiomatic approaches to reconstructing quantum theory~\cite{hrbook,C1_HR2002,C1_HKR2003,C1_RH2011,C1_RH2012,C1_R2014}.  Further, such models have been applied directly to model gravity via the coupling of ensembles of quantum fields to classical spacetimes~\cite{ark,hr2005,hrbook,marcel} (see also section~\ref{sec:grav} below).

We will show in this section that the configuration ensemble approach allows the creation of entanglement between two quantum systems that interact via a classical mediator. Thus the proposals in \cite{bose} and~\cite{marl} cannot unambiguously distinguish between `classical' and `nonclassical' mediators. We first briefly introduce some basic aspects of the configuration ensemble approach below, as it is not widely known.

\subsection{Ensembles on configuration space}

An ensemble on some configuration space $Z$ is described at time $t$ by a probability density $P_t(z)$ on $Z$. Requiring the dynamics of the ensemble to be described by an action principle implies there is a canonically conjugate quantity $S_t(x)$ on configuration space and an ensemble Hamiltonian $H[P,S]$, with corresponding Hamiltonian equations of motion
\beq \label{hameq}
\frac{\partial P}{\partial t} = \frac{\delta H}{\delta S},\qquad
\frac{\partial S}{\partial t} = -\frac{\delta H}{\delta P} .
\eeq
Here $\delta F/\delta f$ denotes the variational derivative of the functional $F[f]$ with respect to function $f$, and reduces to the usual partial derivative for discrete configuration spaces. Observables are represented by suitable functionals of $P$ and $S$, and the Poisson bracket of any two observables $A[P,S]$, $B[P,S]$ is defined in the usual way by
\beq \label{pb}
\{ A,B\} = \int dz \left(\frac{\delta A}{\delta P}\frac{\delta B}{\delta S} - \frac{\delta B}{\delta P} \frac{\delta A}{\delta S} \right) ,
\eeq
with integration replaced by summation over any discrete regions of the configuration space. Thus, the above equations of motion can be rewritten as $\partial P/\partial t=\{P,H\}$ and $\partial S/\partial t=\{S,H\}$, implying that observables evolve via $\partial A/\partial t=\{A,H\}$. Observables, including the Hamiltonian $H[P,S]$, must satisfy certain properties to ensure that probabilities remain positive and normalised~\cite{hrbook,consist}. However, these need not be explicitly considered here.

For example, for a classical configuration space labelled by position $x$, the classical observable $C_f$ corresponding to the phase space function $f(x,k)$ is defined by the functional
\beq \label{cf}
C_f[P,S] := \int dx\, P\,f(x,\nabla_x S) .
\eeq
Note that it is numerically equal to the ensemble average of $f(x,k)$, providing one associates momentum $k=\nabla_x S$ with position $x$.  Evaluating the Poisson bracket of any two classical observables $C_f, C_g$ via Eq.~(\ref{pb}) yields
\beq \label{cb}
\{C_f,C_g\} = C_{\{f,g\}},
\eeq
where $\{f,g\}=\sum_i \left(\frac{\partial f}{\partial x_i}\frac{\partial g}{\partial k_i} - \frac{\partial f}{\partial k_i}\frac{\partial g}{\partial x_i}\right)$ denotes the usual phase space bracket. Thus, the Poisson bracket for classical ensembles is isomorphic to the classical phase space bracket, implying that it generates the standard classical dynamics (and a continuity equation for $P$~\cite{hrbook,consist}).  Thus, Eq.~(\ref{cf}) is an  isomorphism  between  the algebra of observables $C_f$ on configuration space  and  the algebra of observables $f$ on classical phase space.

Similarly, for a quantum configuration space labelled by the possible outcomes $q$ of some complete basis set $\{|q\rangle\}$ of a Hilbert space $\cal H$ (i.e., $\int dq\,|q\rangle\langle q|=\hat \id$, with integration replaced by summation for discrete ranges of $q$), the quantum observable $Q_{\hat M}$ corresponding to the Hermitian operator $\hat M$ is defined by the functional
\beq \label{qm}
Q_{\hat M}[P,S] := \langle \psi|\hat M|\psi\rangle,
\eeq
where $|\psi\rangle\in \h$ is the wave function defined via $\langle q|\psi\rangle=\sqrt{P(q)}e^{iS(q)/\hbar}$. Note that $Q_{\hat M}$ is numerically equal to the ensemble average of $\hat M$ for quantum state $|\psi\rangle$. Further, evaluating the Poisson bracket of any two quantum observables $Q_{\hat M}, Q_{\hat N}$ via Eq.~(\ref{pb}) yields~\cite{hrbook,consist}
\beq \label{qb}
\{ Q_{\hat M}, Q_{\hat N}\} = Q_{[\hat M,\hat N]/(i\hbar)},
\eeq
where $[\hat M,\hat N]$ is the usual commutator. Thus, the Poisson bracket for quantum ensembles is isomorphic to the quantum commutator, implying that it generates the usual Schr\"odinger equation.  Thus, Eq. (\ref{qm})  is an isomorphism between the algebra of observables $Q_{\hat M}$ on configuration space and the algebra of quantum observables $\hat M$.

To describe a hybrid ensemble, comprising two quantum systems $Q$ and $Q'$ and  a classical system $C$ as in figure~\ref{figure}, is straightforward. The joint configuration space is labelled by $(q,q',x)$, and the state of an ensemble on this configuration space is described by a probability density $P(q,q',x)$ and a conjugate quantity $S(q,q',x)$.  Further, the extension of observables on $C$, $Q$ and $Q'$ to the joint configuration space is given by extending the integration in equation~(\ref{cf}) to the joint configuration space, and the wave function in equation~(\ref{qm}) to the hybrid wave function defined via
\beq \label{psi}
\psi(q,q',x):=\sqrt{P(q,q',x)}\,e^{iS(q,q',x)/\hbar} .
\eeq
The Poisson bracket properties (\ref{cb}) and (\ref{qb}) remain unchanged under this extension~\cite{hrbook}. Thus, since the bracket is preserved under Hamiltonian evolution, the algebra of classical observables remains isomorphic to the classical phase space bracket, and the algebra of quantum observables remains isomorphic to the quantum commutator---even under interactions between the classical and quantum components. It should be noted that the hybrid wave function is merely a useful formal construct here, and will not in general evolve linearly~\cite{hr2005,hrbook}.

Finally, we note that while classical systems with discrete configuration spaces can also be described in the configuration ensemble approach~\cite{hrbook}, it suffices, for the purpose of providing counterexamples to the general claim in \cite{bose} and~\cite{marl},  to consider a classical particle with a coordinate $x$ as above.

\subsection{Counterexample: entangling two quantum particles}

We  first consider a simple example in which entanglement can be created via quantum-classical interaction followed by a measurement on the classical system. As noted at the end of section~\ref{sec:koop}, such entanglement creation is not possible for the Koopman-type models considered in \cite{bose} and~\cite{marl}.  While the proposed experiments in these references do not require making such a measurement, this example nevertheless illuminates the strength of the assumptions made in these references.  A more general example, not requiring such a measurement, will be considered further below.

In particular, let $Q$ and $Q'$ be quantum particles with respective positions labelled by $q$ and $q'$, and consider the ensemble Hamiltonian
\begin{align} \label{simple}
H[P,S] &= H_{QC}[P,S] + H_{Q'C}[P,S] \nn \\
&= g_1\int dq\,dq'\,dx \,P\,(\partial_q S) \, x + g_2\int dq\,dq'\,dx \,P\,q'(\partial_x S).
\end{align}
Here $g_1$ and $g_2$ are coupling constants, and we have taken all systems to be one-dimensional for convenience. This choice corresponds to coupling the momentum of $Q$ to the position of $C$, and coupling the position of $Q'$ to the momentum of $C$~\cite{hr2005,hrbook}, as will be seen more clearly below, and thus provides an instance of the scenario in figure~\ref{figure}.
Note  that $H[P,S]$ in equation~(\ref{simple}) corresponds to a pure `interaction' Hamiltonian, to which one would typically add  kinetic energy terms. However, these will be ignored here for convenience, as they are not relevant to the argument. Indeed, taking the coupling constants $g_1$ and $g_2$ to be sufficiently large allows us to ignore any such additional contributions over a short time interval.

It may be shown that the expectation values of all observables of $Q'$ are invariant under the action of the ensemble Hamiltonian $H_{QC}$ alone, and similarly that the expectation values of all observables of $Q$ are invariant under the action of $H_{Q'C}$ alone, as required in the scenario of figure~\ref{figure} (see also chapter~3 of~\cite{hrbook}).
Further, the equations of motion for $P$ and $S$ follow via~(\ref{hameq}) and~(\ref{simple}) as
\beq
\frac{\partial P}{\partial t} = - g_1 x\frac{\partial P}{\partial q}  - g_2 q'\frac{\partial P}{\partial x},\qquad
\frac{\partial S}{\partial t} = - g_1 x\frac{\partial S}{\partial q}  - g_2 q'\frac{\partial S}{\partial x} ,
\eeq
which are easily solved to give
\begin{align}
P_t(q,q',x) &= P_0(q-g_1tx+\half g_1g_2t^2q',q',x-g_2tq'),\\
S_t(q,q',x) &= S_0(q-g_1tx+\half g_1g_2t^2q',q',x-g_2tq') .
\end{align}
Note for $g_2=0$ that the position of Q is displaced by an amount proportional to the position of C, as expected for a position-momentum coupling (both in classical and quantum mechanics). Similarly, for $g_1=0$, the position of $C$ is displaced by an amount proportional to the position of $Q'$.

It immediately follows that the hybrid wave function in equation~(\ref{psi}) evolves linearly in the particular case of  this (rather simple)   ensemble Hamiltonian, with
\beq
\psi_t(q,q',x) = e^{-it(g_1\hat p\hat x+g_2\hat q'\hat k)/\hbar}\,\psi_0(q,q',x) ,
\eeq
where $\hat p$ and $\hat k$ denote the momentum operators $(\hbar/i)\partial_q$ and $(\hbar/i)\partial_x$ conjugate to the position operators $\hat q\equiv q$ and $\hat x\equiv x$. This may easily be checked using the Baker-Campbell-Hausdorff identity $e^{\hat A+\hat B} = e^{\hat A} e^{\hat B} e^{-[\hat A,\hat B]/2}$ for operators $\hat A$ and $\hat B$ that commute with their commutator.  Hence, for this example, the configuration-ensemble dynamics correspond to a formal embedding of the classical particle C into a quantum system and evolving under the Hamiltonian operator $\hat H=g_1\hat p\hat x+g_2\hat q'\hat k$ Note, however, that this correspondence does not extend to treating C itself as fully equivalent to a quantum system, since classical observables as per Eq.~(\ref{cf}) cannot in general be represented by functions of $\hat x$ and $\hat k$. Note also that, unlike the Koopman-like models in section~\ref{sec:koop}, this is not a `diagonal' embedding, as $\hat x$ and $\hat k$ do not commute.  It is this feature that allows the creation of entanglement in the scenario of figure~\ref{figure}.

In particular, suppose that the ensembles are initially independent, with~\cite{hr2005,hrbook}
\beq \label{ind1}
P_0(q,q',x) = P_Q(q) P_{Q'}(q') P_C(x),~~~S_0(q,q',x) = S_Q(q) +S_{Q'}(q') +S_C(x).
\eeq
Note this corresponds to a factorisable hybrid wave function
\beq \label{ind2}
\psi_0(q,q',x)=\psi_Q(q)\, \psi_{Q'}(q')\, \psi_C(x) .
\eeq
It follows that, if a measurement of the classical position $x$ is made after the interaction, with result $x=a$, the statistics of the two quantum particles is described by the quantum wave function
\begin{align}
\psi_{t|a}(q,q') &= K_a \psi_t(q,q',a)
= K_a\psi_Q(q-g_1ta+\half g_1g_2t^2  q' )\,\psi_{Q'}(q')\,\psi_C(a-g_2tq') 
\end{align}
 on $\h_Q\otimes\h_{Q'}$,  where $K_a$ is a normalisation constant.  Clearly, this wave function does not factorise, and hence is entangled,  providing a counterexample  to the predictions of the proposals in \cite{bose} and~\cite{marl} (see the last paragraph of section~\ref{sec:koop}).

It would be of interest, for the above example, to also consider whether the mixture described by the  quantum  density operator
\beq
\hat\rho_{QQ'|C}(t) := \int_a p(a) |\psi_{t|a}\rangle\langle\psi_{t|a}|
\eeq
is entangled under suitable initial conditions, where $p(a)=  |K_a|^{-2} $ is the probability of measuring $x=a$, and $\langle q,q'|\psi_{t|a}\rangle:=\psi_{t|a}(q,q')$. This could perhaps be investigated for the case of Gaussian states.  However, a more general example is given below, for which it is clear that the corresponding density operator can indeed be entangled after the interaction.

\subsection{Counterexample: entangling two general\\ quantum systems}

We now consider general quantum systems $Q$ and $Q'$, rather than one-dimensional particles, and the ensemble Hamiltonian
\begin{align}
H'[P,S] &= H'_{QC}[P,S] + H'_{Q'C}[P,S] \nn \\
&= \int dq\,dq'\,dx \, \psi^*(q,q'x)\, x\hat M \,\psi(q,q'x)\nn\\
&~~ + \half \int dq\,dq'\,dx \, \psi^*(q,q'x)\left[ \frac{\partial S(q,q',x)}{\partial x}\hat N +\hat N\frac{\partial S(q,q',x)}{\partial x} \right]\psi(q,q'x) .
\end{align}
Here $\psi(q,q'x)$ is the hybrid wave function in~(\ref{psi}), and $\hat M$ and $\hat N$ are operators acting on $\h_Q$ and $\h_{Q'}$ respectively. As before, any additional `noninteraction' contributions to the Hamiltonian have been ignored for convenience. We note that this ensemble Hamiltonian satisfies the required consistency constraints for ensuring conservation and positivity of probability~\cite{hrbook, consist}.

It is straightforward to check that $H'[P,S]=\langle\psi|\hat x \hat M+ \hat k \hat N|\psi\rangle$, with $\hat x$ and $\hat k$ defined as per the previous example. Hence, in this particular case, the corresponding equations of motion for $P$ and $S$ can be rewritten as a linear Schr\"odinger equation
\beq
i\hbar\frac{d}{dt}|\psi\rangle = (\hat M\otimes \id_{Q'}\otimes \hat x + \hat \id_Q\otimes \hat N \otimes \hat k)|\psi\rangle
\eeq
for the hybrid wave function, again allowing the dynamics (although not all classical observables) to be embedded in a fully quantum system. Assuming initially independent ensembles as per~(\ref{ind1}) and~(\ref{ind2}), and again using the Baker-Campbell-Hausdorff relation, it follows that the evolution is described by (dropping explicit tensor products)
\beq
|\psi_t\rangle = e^{-it(\hat x\hat M+\hat k\hat N)/\hbar}|\psi_0\rangle = e^{-it\hat x\hat M/\hbar}e^{-it\hat k\hat N/\hbar}e^{it^2\hat M\hat N/(2\hbar)}|\psi_Q\rangle|\psi_{Q'}\rangle|\psi_C\rangle .
\eeq
The rightmost unitary operator clearly entangles $Q$ and $Q'$, thus providing a general counterexample to the claims made in~\cite{bose, marl}.

\section{A `gravitational' counterexample?}
\label{sec:grav}

The general scenario considered in \cite{bose} and~\cite{marl} and figure~1 is not restricted by any known physical properties of gravity or general relativity (such as, e.g., a inverse-square law or covariance under coordinate transformations).  It is therefore sufficient to find  counterexamples that are similarly not restricted in this way, as was done in the previous section.  However, it is also of interest to investigate whether a counterexample can still be constructed under the imposition of physically-motivated constraints on gravitational interactions. In particular, if the creation of entanglement between $Q$ and $Q'$ becomes impossible under a given set of such constraints, then the proposals in \cite{bose} and~\cite{marl} will have the merit of providing an unambiguous witness of nonclassical gravitational effects relative to these constraints.

Given the above, it is  relevant that a configuration-ensemble model for the coupling of quantum fields to classical spacetime has been previously put forward~\cite{hr2005}.  This model is consistent with the classical description of gravity as per general relativity, and some initial investigations of its properties have been made for minisuperspace and midisuperspace models~\cite{hr2005,hrbook,marcel}. A detailed description of how  quantum matter fields couple to  classical spacetime in the configuration-space formalism is given in~\cite{hrbook}. We provide a brief overview in the Appendix to this paper, focusing on the particular case of the coupling of quantum scalar fields to the metric field; see in particular equations~(\ref{Appendix-hh})-(\ref{Appendix-cehj}). While we are unable at the time of writing to determine whether this particular model is able to generate entanglement between two quantum fields, we can point to some suggestive nonclassical features.

One aspect that we would like to emphasise here, which is already apparent from the form of these equations and that of the ensemble Hamiltonian from which they are derived, equation~(\ref{Appendix-ehqphih}), is that there is no sense in which the interaction between each of the quantized fields and the classical gravitational field can be `switched off' -- matter bends space and space curves matter, and a change in one component will drive a change in the other component  (and is why the model does not lead to nonlocal signalling issues~\cite{hrbook,savage}).  This corresponds to the direct multiplicative coupling of the metric tensor to the quantized fields in the ensemble Hamiltonian of equation~(\ref{Appendix-ehqphih}), which is a fundamental feature of gravitational interactions. Thus it would not be unreasonable to expect entanglement to be generated between the various quantum fields, similarly to the examples of the previous section, resulting from their interaction with their common classical gravitational field. 

Furthermore, we note that the case of a classical CGHS black hole and a quantized scalar field can be solved \cite{hrbook}, with Hawking radiation being predicted for a collapsing space-time geometry. This is another indication that a classical gravitational field can have non-trivial effects on quantized matter fields in hybrid models.

\section{Discussion}
\label{sec:disc}

It has been shown that the validity of the claim relied on by Bose {\it et al.}~\cite{bose} and Marletto and Vedral~\cite{marl}, that the scenario of figure~\ref{figure} cannot create entanglement between two quantum systems, is entirely dependent on the model of quantum-classical interactions.  As shown here, for example, the claim is valid for both Koopman and mean-field models but not for general configuration-ensemble models.

It follows that an observed generation of entanglement via the experimental proposals made in~\cite{bose} and~\cite{marl} would not provide a definitive test of the nonclassical nature of gravity, unless further physical constraints are imposed which rule out all possible classical explanations. The constraints corresponding to the specific model in section~\ref{sec:grav} are of interest in this regard, with further investigation needed,  particularly in the weak-field limit,  to determine whether it is able to create entanglement between quantum fields.

We close by briefly remarking on some subtleties related to the description of quantum-classical interactions.  First, as noted in section~\ref{sec:overview}, there are a number of no-go results that rule out many possible interaction models as inconsistent~\cite{terno,diosi,salcedo,C9_CS1999,sahoo}.  For example, the Koopman dynamics of a coupled classical and quantum oscillator leads to unphysical amplitude oscillations that are inconsistent with the classical limit~\cite{terno}.  Moreover, even minimally-consistent models can have a number of unusual properties~\cite{hrbook,salcedocrit,radon2}.

Second, while the notion of `entanglement' of two quantum systems $Q$ and $Q'$ is not problematic for Koopman and mean-field models, since a corresponding quantum state is always well-defined in these models, this is not the case for more general models. For the counterexamples in section~\ref{sec:config}, this was dealt with by demonstrating quantum entanglement relative to the configuration of the classical system. This raises the question of how entanglement might be defined more broadly.  A preliminary discussion of `entanglement' for general configuration ensembles has been given in chapter~3 of~\cite{hrbook}.

Third and finally, we note that is not strictly correct to refer to the proposals by Bose {\it et al.} and by Marletto and Vedral as providing a test or witness of {\it quantum} gravity, as is done in the titles and elsewhere of \cite{bose} and~\cite{marl}. The best that can be said is that they can provide, under suitable constraints, a test or witness of {\it nonclassical} gravity.  Whether such nonclassicality is quantum in nature, or otherwise, remains an open question.

{\flushleft{\bf Acknowledgment ---} MH is supported by the ARC Centre of Excellence CE110001027.}

\section*{Appendix: Coupling of scalar quantum fields to classical gravity in the formalism of ensembles on configuration space}

We summarize the equations that describe the coupling of scalar quantum fields to classical gravity. We proceed by steps: we first define a classical configuration space ensemble for pure gravity and then consider the addition of quantum fields. A detailed description of the formalism is given in~\cite{hrbook}.

The most direct way of introducing a classical configuration space ensemble for gravity is to start from the Einstein-Hamilton-Jacobi equation, which in the metric representation takes the form \cite{MTW73}
\begin{equation}\label{Appendix-ehj}
H^C_h =\kappa G_{ijkl}\frac{\delta S}{\delta h_{ij}}\frac{\delta S}{
\delta h_{kl}}-\frac{1}{\kappa }\sqrt{h}\left( R-2\lambda \right) =0 .
\end{equation}
Here $\kappa =16\pi $ (in units where the speed of light $c$ and the gravitational constant $G$ are equal to one, i.e., $c=G=1$); $G_{ijkl}=\left( 2h\right) ^{-1/2}\left( h_{ik}h_{jl}+h_{il}h_{jk}-h_{ij}h_{kl}\right) $ is the
DeWitt supermetric \cite{MTW73}, $h_{ij}$ is the spatial part of the metric tensor, with determinant $h$; $R$\ is the curvature scalar; $\lambda $ is the cosmological constant; and $D_{j}$ is the spatial covariant derivative.The functional $S$ is assumed to be invariant under the gauge group of spatial coordinate transformations, which is equivalent to satisfying the momentum constraints of the canonical formulation of general relativity. Equation~(\ref{Appendix-ehj}) corresponds to an infinity of constraints, one at each point. It is possible to introduce an alternative viewpoint \cite{K93} in which equation~(\ref{Appendix-ehj}) is regarded as an equation to be integrated with respect to a ``test function'' in which case we are dealing with one equation for each choice of lapse function $N$,
\begin{equation} \label{Appendix-hg}
\int d^{3}x\,\,N H^C_h=0;
\end{equation}
i.e., for each choice of foliation. Such an alternative viewpoint is extremely useful: although it may be impossible to find the general solution (which requires solving the Einstein-Hamilton-Jacobi equation for all choices of lapse functions), it may be possible to find particular solutions for specific choices \cite{K93}.

Taking Eq. (\ref{Appendix-hg}) into consideration, we define the ensemble Hamiltonian for vacuum gravity according to
\begin{equation} \label{Appendix-eehj}
{\cal H}^C_h=\int d^{3}x N \int Dh\,P\,\,H^C_h,
\end{equation}
where $Dh$ is an appropriate measure over the space of metrics (technical issues are discussed in~\cite{hrbook}). The functional $P$ is also assumed, like $S$, to be invariant under the gauge group of spatial coordinate transformations. The corresponding equations have the form
\begin{equation}
\frac{\partial P}{\partial t}=\frac{\Delta {\cal H}^C_h}{\Delta S},\quad
\frac{\partial S}{\partial t}=-\frac{\Delta {\cal H}^C_h}{\Delta P}, \nonumber
\end{equation}
where $\Delta /\Delta F$ denotes the variational derivative with respect to a functional $F$ (see Appendix A of~\cite{hrbook}). Assuming the constraints $\frac{\partial S}{\partial t}=$ $\frac{\partial P}{\partial t}=0$, these equations lead to equation~(\ref{Appendix-hg}), as required, and to a continuity equation,
\begin{equation} \label{Appendix-cehj0}
\int d^{3}x \,N \frac{\delta }{\delta h_{ij}}\left( P\,G_{ijkl}\frac{\delta S}{
\delta h_{kl}}\right) =0.
\end{equation}
These two equations, Eqs. (\ref{Appendix-hg}) and (\ref{Appendix-cehj0}), define the evolution of an ensemble of classical spacetimes on configuration space for the case of vacuum gravity.

A hybrid system, where $n$ quantum scalar fields $\phi_a$ couple to the classical metric $h_{kl}$, requires a generalization of equation~(\ref{Appendix-eehj}) in which
\begin{equation} \label{Appendix-ehqphih}
{\cal H}_{\phi h}=\int d^{3}x \,N \int Dh D\phi \,P  \left[ {H}^C_{\phi h}+F_{\phi }
\right] ,
\end{equation}
where
\begin{equation}
{H}^C_{\phi h}= {H}^C_{h}+
\sum_{a=1}^n \left\{ \frac{1}{2\sqrt{h}}\left( \frac{\delta S
}{\delta \phi_a }\right) ^{2}+\sqrt{h}\left[ \frac{1}{2}h^{ij}\frac{\partial
\phi_a }{\partial x^{i}}\frac{\partial \phi_a }{\partial x^{j}}+V\left( \phi_a
\right) \right] \right\} \nonumber
\end{equation}
and
\begin{equation}
F_{\phi }=\sum_{a=1}^n \left\{ \frac{\hbar ^{2}}{4}\frac{1}{2\sqrt{h}}\left( \frac{\delta \log P
}{\delta \phi_a }\right) ^{2} \right\}. \nonumber
\end{equation}
To interpret the terms that appear in Eq. (\ref{Appendix-ehqphih}), note that ${H}^C_{\phi h}=0$ is the Einstein-Hamilton-Jacobi equation for gravity with $n$ classical scalar fields and that $F_{\phi }$ is a non-classical term which accounts for the quantum nature of the fields. Assuming again the constraints $\frac{\partial S}{\partial t}=$ $\frac{\partial P}{\partial t}=0$, the corresponding equations are given by
\begin{equation} \label{Appendix-hh}
\int d^{3}x\, N \, \left[ \mathcal{H}^C_{\phi h}
-\sum_{a=1}^n \left\{ \frac{\hbar ^{2}}{2\sqrt{h}}
\left( \frac{1}{A}\frac{\delta ^{2}A}{\delta \phi_a ^{2}}\right) \right\} \right]=0,
\end{equation}
where $A\equiv \sqrt{P}$, and a continuity equation of the form
\begin{equation} \label{Appendix-cehj}
\int d^{3}x \,N \left[ \frac{\delta }{\delta h_{ij}}\left( P\,G_{ijkl}\frac{\delta S}{
\delta h_{kl}}\right) +  \sum_{a=1}^n \left\{ \frac{1}{\sqrt{h}} \frac{\delta }{\delta \phi_{a}}\left( P \frac{\delta S}{
\delta \phi_{a}}\right) \right\} \right] =0.
\end{equation}.

\end{document}